\documentclass{iopart}  
\usepackage{graphics}
\usepackage{epsfig}

 \newcommand{\be}[1]{\begin{equation}\label{#1}}
\newcommand{\ee}{\end{equation}}   
  
\newcommand{\bea}{\begin{eqnarray}}
\newcommand{\eea}{\end{eqnarray}} 
\newcommand{\eq}[1]{(\ref{#1})}

\begin{document}  
\jl 2
\letter{A quasi classical approach to fully differential ionization cross 
	sections}
\author{Tiham\'{e}r Geyer\dag \ and Jan M Rost\ddag}
\address{\dag-- Theoretical Quantum Dynamics --
Fakult\"at f\"ur Physik, Universit\"at Freiburg,
Hermann--Herder--Str.  3, D--79104 Freiburg,
Germany}
\address{\ddag Max--Planck--Institute for the Physics of Complex Systems, 
N\"othnitzer Str. 38, D--01187 Dresden,
Germany}
\date{\today}	
\begin{abstract}
\noindent A classical approximation to time dependent quantum mechanical 
scattering in the M\o{}ller formalism is presented.  Numerically, our 
approach is similar to a standard Classical--Trajectory--Monte--Carlo 
calculation.  Conceptually, however, our formulation allows one to release 
the restriction to stationary initial distributions.  This is achieved 
by a classical forward--backward propagation technique.  As a first 
application and for comparison with experiment we present fully 
differential cross sections for electron impact ionization of atomic 
hydrogen in the Erhardt geometry.
 \end{abstract} 
\vspace{5mm}

Classical models and approximations are frequently used for atomic and 
molecular problems, despite their inherent quantum nature. One of the 
reasons is that our intuitive understanding is mainly based on 
classical terms and pictures by which we are surrounded in every day 
life. Another reason for not doing (fully) quantum mechanical 
calculations is the complexity of a problem: fully differential cross sections 
in higher dimensional atomic
systems, e.g., often require a numerical effort still beyond present computing power. 
Only recently the quantum mechanical Coulomb three--body scattering 
problem was solved numerically \cite{BRA99,RES99}.

On the other hand, a remarkably successful classical approach to 
collisional atomic problems has been developed over the years, the so 
called Classical--Trajectory--Monte--Carlo method (CTMC). It was 
introduced as a purely classical model based on a ``planetary atom" 
with a major axis of two meters (!) \cite{ABR66}. This model has 
produced reasonable results for total or energy differential 
ionization cross sections on atomic hydrogen \cite{CHydxx, OLS77} and 
for other few--body Coulomb collision processes \cite{SCH92}.

Attempts to reduce the limitation
of this classical model aimed at changing the description of the 
initial state for the 
hydrogen atom from a microcanonical distribution to one, that is 
closer
to the quantum density \cite{HAR83,EIC81}. Another idea was to 
introduce additional ad hoc stabilisation potentials in order to be 
able to treat multi--electron targets \cite{ZAJ86}. However, all 
attempts took as a starting point not the quantum problem but the 
previously formulated classical model. Hence, the proposed amendments 
were accompanied by inconsistencies or the need of "fit parameters" 
determined from cross sections. In the end, one must say that it is up 
to now not possible to describe higher differential cross sections or 
targets with more than one active electron consistently in a classical 
collision framework.

To achieve progress in this situation we decided to go one step 
backwards and start with a time dependent quantum mechanical 
scattering formalism. By following the approximations which lead to 
the classical description, i.e. the CTMC method, we can identify the 
source and nature of deviations between the classical and the quantum
result which serve as a guide to improve the classical description to 
a quasi--classical approximation.

We divide the problem into three logically separate steps: (1) 
preparation of the initial state before the collision, (2) propagation 
in time, and (3), extraction of the cross section. For a consistent quasiclassical 
picture each of these steps has to be approximated in the same way. To 
keep the derivation transparent, we will concentrate on electron 
impact ionization of one active target electron in the following.

\smallskip Step (1): The initial wave function for the collision 
problem is translated first into a quantum phase space distribution by 
the Wigner transformation \cite{WIG32}. The resulting Wigner 
distribution is reduced to a classical distribution $w(p,q)$ which can 
be propagated classically in phase space by taking the usual limit 
$\hbar\to 0$. The difference to an a priori classical approach is the 
use and interpretation of negative parts of the distribution: Viewed 
as the $\hbar \to 0 $ limit of a quantum problem they contribute to 
the observables in the same way as the positive parts since they do 
not need to be interpreted as weights for real paths of classical 
particles. Yet, there arise additional problems when using this type 
of general initial phase space distributions in the usual classical 
framework:
Most of them are not stationary under classical propagation, their 
Poisson bracket with the Hamilton function does not vanish, 
$\{H,w\}\ne 0$. Hence, the initial target distribution will look very 
different at the time the projectile has approached and the collision 
actually happens.

\smallskip

Step (2): The formulation of the propagation is crucial since it must 
resolve the problem of the non--stationary classical initial 
distribution, as described above. Traditionally, the time dependent 
scattering is described by calculating the transition amplitude 
between initial and final state through the S--matrix, which is in turn 
related to the t--matrix
describing directly the cross section, see, e.g., \cite{TAY72}. In a simplified 
version where the asymptotic initial and final states are eigenstates 
of the asymptotic Hamiltonians $H^{(i)}_0$ and $H^{(f)}_0$ one 
normally writes for the transition amplitude
\be{S-simple}
	S_{fi}= \lim_{t\to\infty}\langle f|U(t)|i\rangle
\ee
where $U(t) = \exp[-iHt]$ denotes propagation with the full 
Hamiltonian. By a Wigner transform the quantum time evolution operator 
$U(t)$ can be directly transformed with the help of the quantum 
Liouville operator $\mathcal{L}_q$, which reduces to the classical 
Liouville operator $\mathcal{L}_c$ in the limit $\hbar\to 0$ \cite{HEL76}. 
The latter describes the evolution of a phase space distribution 
$w(r,p,t)$ according to the Poisson bracket
\be{lio}
	\partial_{t}w = \{H,w\} \equiv -i\mathcal{L}_c w
\ee
in  analogy to the quantum evolution of the density matrix $\rho$ 
generated by the commutator,
\be{lio2}
	\partial_{t}\rho= -i[H,\rho].
\ee
Hence, we could directly use the translation of \eq{S-simple} to classical 
mechanics via the Liouville operator.  In connection with the  microcanonical
initial state distribution this is indeed equivalent to the  CTMC
formulation \cite{KEL93}.  
However, using non--stationary initial state  distributions is
inconsistent with the reduced quantum description of  \eq{S-simple} which
relies on the fact that the asymptotic states are  eigenstates of $U_{0}$ and
therefore stationary.  Instead we have to go  back to the full scattering
formulation
\be{S-mat}
	S_{fi}= \langle f|\Omega^\dagger_{-}\Omega_{+}|i\rangle,
\ee
where
\be{moel}
	\Omega_{\mp} = \lim_{t \to \pm \infty} U^{\dagger}(t)\,U_0(t)
\ee
are the M\o{}ller operators.  The meaning of $\Omega_{+}$, e.g., is to 
propagate backwards with $U_{0}(t)$ using the asymptotic Hamiltonian 
$H_{0}^{i}$ without the projectile--target interaction and then forward 
again under the full Hamiltonian with $U(t)$.  Again, with the help of the 
Liouville operator we can translate the M\o{}ller operators to their 
classical analogue, thereby obtaining a  prescription how to propagate a 
non--stationary initial phase space distributions $w_i(\gamma)$, where
$\gamma = (\vec p_1,\vec q_1,\vec p_2,\vec q_2)$ is a point in the 
12-dimensional phase space:
\be{moelclas}
	w_f = \lim_{t \to +\infty}\lim_{t' \to -\infty} 
	e^{-i\mathcal L^{f}_ct}e^{i\mathcal L_c(t-t')}
	e^{i\mathcal L^{i}_ct'}w_i 
	\equiv {\cal K}w_i\,.
\ee

The difference to \eq{S-simple} are the explicit propagations under 
$\mathcal L^{f}_c$ and $\mathcal L^{i}_c$ in the initial and final 
channel (which need not be the same). The meaning of \eq{moelclas} becomes
very transparent if we insert a discretized distribution, which is used in the 
actual calculations, $w_i(\gamma) = \sum_n w_n\delta^{12}(\gamma-\gamma_n^{i})$.
The final distribution reads
\be{moelclas2}
w_f(\gamma) = {\cal K}w_i = \sum_nw_n\delta^{12}(\gamma-\gamma_n^{f})
\ee
where each phase space point $\gamma_n^{f}$ emerges from $\gamma_n^{i}$ through
solving successively Hamilton's equations, first with $H_0^{i}$, then with $H$, and
eventually with $H_0^f$.
With this propagation scheme  a non--stationary 
initial distribution will spread when being propagated backwards with 
the asymptotic $\mathcal L_{c}^{i}$. However, it will be refocused 
under the following forward propagation with $\mathcal L_{c}$. Hence, 
when the actual collision happens for $t\approx 0$ the original target 
distribution is restored, slightly polarized by the approaching 
projectile.

Hence, there is no more need for the initial distribution to be 
classically stationary. We are able to use any phase space 
distribution as a target in our quasi classical approach. This also 
includes unstable multi--electron targets, e.g., classical helium.

\smallskip

Step (3): Before we come to the actual evaluation we have to formulate the
cross section such that it can make full use of the non-stationary initial
phase space distribution $w_i(\vec p_1,\vec q_1)$, where ``1'' refers to the
target electron. Without modification the total energy $E$ of the final state
forces by  energy conservation for each classical trajectory  only those parts
of the initial phase space distribution  to contribute to the cross section
which have the same energy $E$. However, this would bring us essentially back
to the microcanonical description. In order to make the entire non-stationary
initial state distribution ``visible'' to the collision process,  we  use the
energy transfer $\bar E_1 = E^{(f)}_1 - E^{(i)}_1$  to the target electron
rather than its energy $E^{(f)}_1$ itself as a differential measure. Of course,
as long as the initial state is on the energy shell with well defined energy $E
= E^{(i)}_1 + E^{(i)}_2$ the new definition coincides with the usual
expression for the cross section,
\be{CrossSec}
\left. \frac{d^5\sigma}{d\Omega_1d\Omega_2dE_1}\right |_{E} =
\left. \frac{d^5\sigma}{d\Omega_1d\Omega_2d\bar E_1}\right |_{E}\,,
\ee
where $d\Omega_i$ are the  differentials for the solid angles of the
two electrons, respectively.

To extract this cross section  we have to  evaluate the phase space 
integral 
\be{phaseSpaceInt}
\noindent 
\frac{d^5\sigma}{d\Omega_1d\Omega_2d\bar E_1}=
	 \int \!\! dx_2\,dy_2\,d\vec p_1\,d\vec q_1\,
	 \prod_{i =1}^2\delta(\Omega^{(f)}_i-\Omega_i) 
	 \delta(\bar E^{(f)}_1-\bar E_1) w_i\,,
\ee
where the integration is over the initial state variables, namely  the impact
parameter area $dx_2\, dy_2$ and the phase space of the (bound) target electron
$d\vec p_1 d\vec q_1$, with  initial distribution  $w_i(\vec p_1,\vec q_1,
x_2,y_2)$. The propagated angles $\Omega^{f}_i$ of the electrons as well as the
energy transfer $\bar E^f_1$ have to coincide with the desired values
$\Omega_i$ and $\bar E_1$ to contribute to the cross section \eq{phaseSpaceInt}
which is a generalization of the one derived in \cite{Ros98}, e.g., where the
initial bound state was assumed to live on a torus, i.e., $w_i(\vec p_1,\vec
q_1) = \delta (\vec I(\vec p_1,\vec q_1)-\vec I_0)$ with a well defined
multidimensional action $\vec I_0$.

Finally, we have to respect the Pauli principle for the two identical
electrons. Formally, this can be done easily in the Wigner transform
for the two ionized electrons in the final state. In the limit 
$\hbar
= 0$ one is left with the usual classical symmetrization, i.e., an
interchange of indices.  To keep the notation simple we have omitted
symmetrization in the outlined derivation, however, it is included in
the actual computation which is carried out by applying standard CTMC
techniques to evaluate \eq{phaseSpaceInt}.


\epsfxsize=3in
\epsfysize=4.5in
\begin{figure}[htb]
	\centerline{\epsfbox{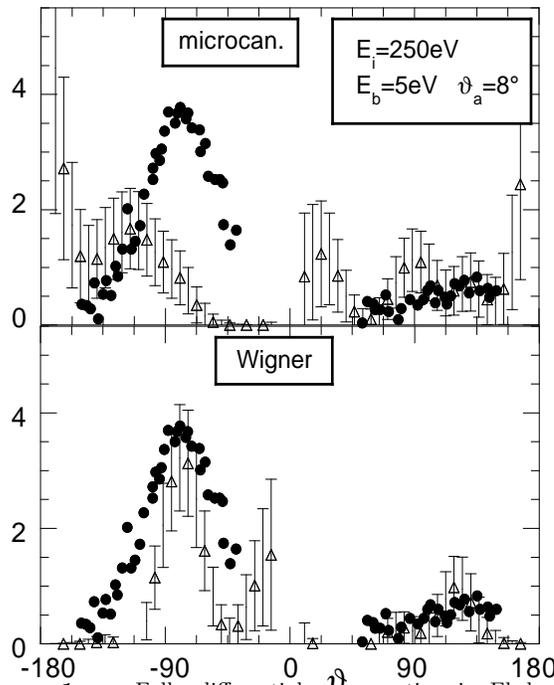}}
	\vspace{-2.8cm} 
	\caption{
	\label{fig:TDCS250}
	Fully differential cross section in 
	Ehrhardt geometry: $E_{in}$= 250eV, $E_b$= 5eV and $\theta_a$= 
	8$^0$: comparison of measurements of Erhardt \etal to quasiclassical 
	calculations with (a), the microcanonical 
	distribution (standard CTMC), and (b) with the Wigner 
	distribution. Error bars 
	indicate the statistical error of the Monte Carlo calculations.
	The theoretical data  has been scaled by about 20\% to reproduce the correct 
	total ionization cross section at 250 eV. For the negative parts in the Wigner cross 
	section see text.}
\end{figure}

\epsfxsize=3in
\epsfysize=4.5in
\begin{figure}[htb]
	\centerline{\epsfbox{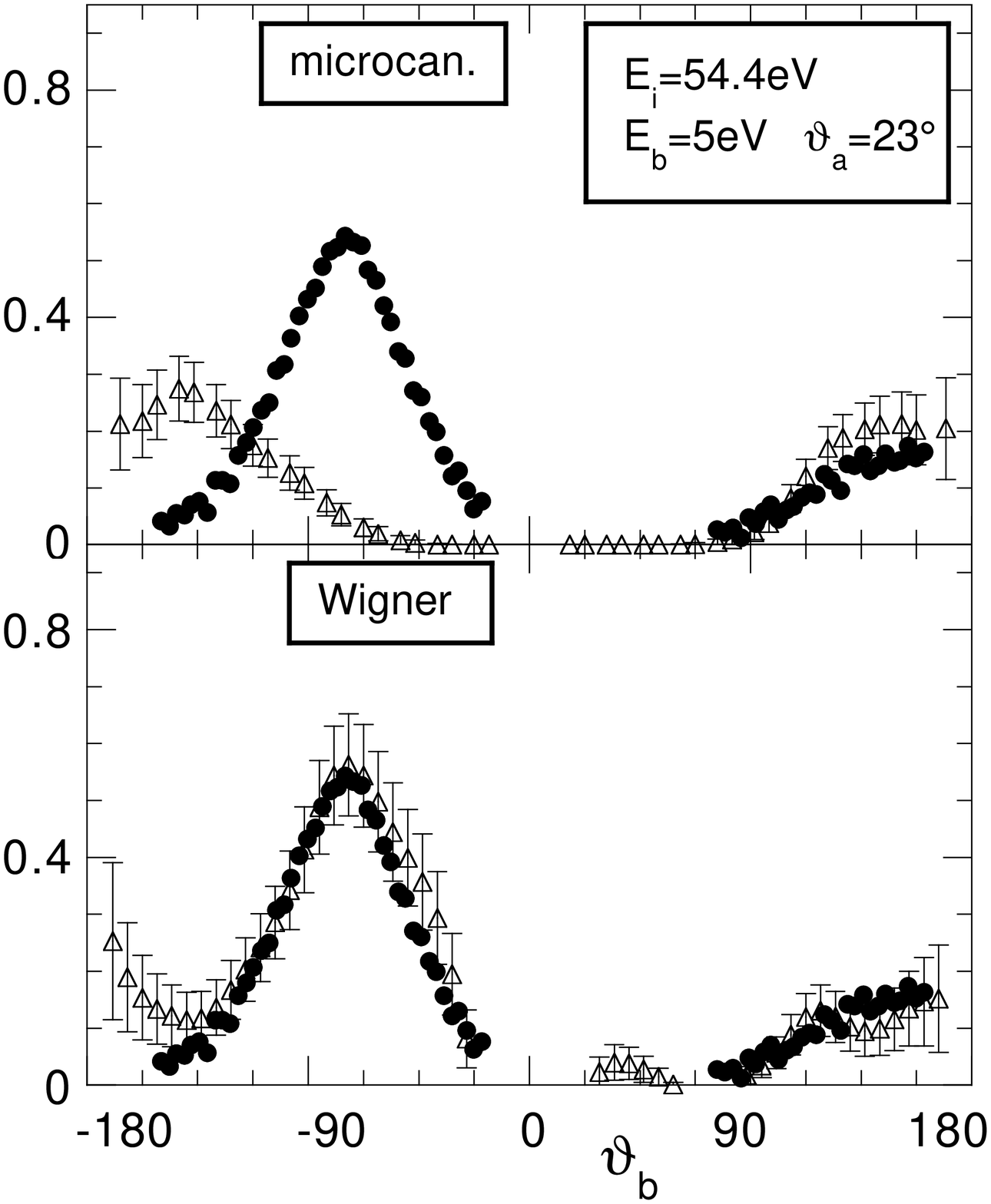}}
	\vspace{-2.8cm} 
 	\caption{Same as in Figure 1, but for 
	$E_{in}$= 54.4eV, $E_b$= 5eV and $\theta_a$= 
	23$^0$ and 
	measurements (circles) by R\"oder \etal \protect\cite{ROD96}. }
	\label{fig:TDCS54}
\end{figure}

As a first application we discuss fully differential ionization cross
sections  for atomic hydrogen comparing three data sets: a calculation
with the standard microcanonical distribution (CTMC), one with the
non--stationary Wigner distribution in our quasi classical  framework
and experimental data at  impact energies of 250eV 
(fig.~\ref{fig:TDCS250}) and 54.4eV (fig.~\ref{fig:TDCS54}), 
respectively.

For each impact energy about $10^8$ trajectories have been calculated. 
The cross section at 250 eV for the Wigner distribution still exhibits 
negative parts, indicating that this cross section is not yet fully 
converged. This is not surprising if one takes into account that the 
fraction of phase space of the final state is so small with the chosen 
bin sizes for energies and angles, that only between 100 and 300 
events finally contribute to the shown cross sections. However, a 
considerable advantage of the present method is that a sampling of 
$10^8$ trajectories contains the complete scattering information, not 
just one specific differential cross section.

The figures show, that the microcanonical distribution, i.e. standard 
CTMC, is not able to reproduce the binary peak \cite{BRI89}, whereas
with the  Wigner distribution it is reproduced fairly well for 250eV
and rather  well for 54.4eV impact energy. Keeping in mind that in
contrast to the  microcanonical distribution the Wigner distribution
has the correct  probability densities in momentum {\it and}
configuration space, one  can conclude, that at least for energies
between $50$ and $250$ eV the  differential cross sections "image" the
initial {\it phase space}  distribution.
The present approach is still a classical approximation and cannot 
reproduce quantum effects. Therefore, features in the cross section, 
for which coherence is crucial are represented purely or not at all.

To summarize, we have shown, that a consequent classical approximation 
to a quantum system can give much better results compared to those 
from an a priori classical model, though both approaches are realized 
numerically in almost the same way. However, the main difference is 
conceptual: in the usual classical limit each individual trajectory 
represents that of an electron obeying the classical equations of 
motion, whereas in our classical approximation only the entire phase 
space distribution is meaningful and individual trajectories are only 
discretized points of the distribution evolving in time. Hence, there 
is no problem to deal with "negative probabilities" in the initial 
distribution, since we regard them not as probabilities but only as 
weights of the integration which, of course, may be negative.

However, the use of non-stationary distributions like the Wigner distribution as
an initial state implies additional difficulties for  a scattering description
which we have overcome by using a forward-backward propagation scheme akin to
the quantum M\o{}ller formalism and by a reformulation of the energy
differential cross section in terms of the energy transfer during the ionization
process. With these modifications all the tools of the standard CTMC technique
can be applied straightforwardly. 

Moreover, our approach can be in principle generalized to 
multi--electron targets since we generate our initial phase space 
distribution from a quantum wave function and we know how to deal with 
non--stationary initial distributions.

Financial support by the Deutsche Forschungsgemeinschaft within the SFB 276
at the University Freiburg is gratefully acknowledged. 


\end{document}